
 %
\input harvmac.tex  
 %
\catcode`@=11
\def\rlx{\relax\leavevmode}                  
 %
 %
 %
\font\tenmib=cmmib10
\font\sevenmib=cmmib10 at 7pt 
\font\fivemib=cmmib10 at 5pt  
\font\tenbsy=cmbsy10
\font\sevenbsy=cmbsy10 at 7pt 
\font\fivebsy=cmbsy10 at 5pt  
\def\BMfont{\textfont0\tenbf \scriptfont0\sevenbf
                              \scriptscriptfont0\fivebf
            \textfont1\tenmib \scriptfont1\sevenmib
                               \scriptscriptfont1\fivemib
            \textfont2\tenbsy \scriptfont2\sevenbsy
                               \scriptscriptfont2\fivebsy}
\def\BM#1{\rlx\ifmmode\mathchoice
                      {\hbox{$\BMfont#1$}}
                      {\hbox{$\BMfont#1$}}
                      {\hbox{$\scriptstyle\BMfont#1$}}
                      {\hbox{$\scriptscriptstyle\BMfont#1$}}
                 \else{$\BMfont#1$}\fi}
 %
 %
 %
 %
\def\inbar{\vrule height1.5ex width.4pt depth0pt}
\def\sinbar{\vrule height1ex width.35pt depth0pt}
\def\ssinbar{\vrule height.7ex width.3pt depth0pt}
\font\cmss=cmss10
\font\cmsss=cmss10 at 7pt
\def\ZZ{\rlx\leavevmode
             \ifmmode\mathchoice
                    {\hbox{\cmss Z\kern-.4em Z}}
                    {\hbox{\cmss Z\kern-.4em Z}}
                    {\lower.9pt\hbox{\cmsss Z\kern-.36em Z}}
                    {\lower1.2pt\hbox{\cmsss Z\kern-.36em Z}}
               \else{\cmss Z\kern-.4em Z}\fi}
\def\Ik{\rlx{\rm I\kern-.18em k}}  
\def\IC{\rlx\leavevmode
             \ifmmode\mathchoice
                    {\hbox{\kern.33em\inbar\kern-.3em{\rm C}}}
                    {\hbox{\kern.33em\inbar\kern-.3em{\rm C}}}
                    {\hbox{\kern.28em\sinbar\kern-.25em{\sevenrm C}}}
                    {\hbox{\kern.25em\ssinbar\kern-.22em{\fiverm C}}}
             \else{\hbox{\kern.3em\inbar\kern-.3em{\rm C}}}\fi}
\def\IP{\rlx{\rm I\kern-.18em P}}
\def\IR{\rlx{\rm I\kern-.18em R}}
\def\Ione{\rlx{\rm 1\kern-2.7pt l}}
 %
 %

 %

\def\intem#1{\par\leavevmode%
              \llap{\hbox to\parindent{\hss{#1}\hfill~}}\ignorespaces}
 %


 %
\newskip\humongous \humongous=0pt plus 1000pt minus 1000pt   
\def\caja{\mathsurround=0pt}
\newif\ifdtup
 %
\def\eqalign#1{\,\vcenter{\openup2\jot \caja
     \ialign{\strut \hfil$\displaystyle{##}$&$
      \displaystyle{{}##}$\hfil\crcr#1\crcr}}\,}
 %

 %
\def\panorama{\global\dtuptrue \openup2\jot \caja
     \everycr{\noalign{\ifdtup \global\dtupfalse
      \vskip-\lineskiplimit \vskip\normallineskiplimit
      \else \penalty\interdisplaylinepenalty \fi}}}
 %
\def\eqalignno#1{\panorama \tabskip=\humongous
     \halign to\displaywidth{\hfil$\displaystyle{##}$
      \tabskip=0pt&$\displaystyle{{}##}$\hfil
       \tabskip=\humongous&\llap{$##$}\tabskip=0pt\crcr#1\crcr}}
 %

 %

 %

 %
 %
 %
 %
\let\ii=\i          
\def\,{\hskip1.5pt}           
 %
\let\a=\alpha
\let\b=\beta
\let\c=\chi
\let\d=\delta       \let\vd=\partial             \let\D=\Delta
\let\e=\epsilon     
\let\f=\phi                       \let\F=\Phi
\let\g=\gamma                                    \let\G=\Gamma
\let\h=\eta
\let\i=\iota
\let\j=\psi                                      \let\J=\Psi
\let\k=\kappa
\let\l=\lambda                                   
\let\m=\mu
\let\n=\nu
\let\p=\pi          \let\vp=\varpi               
                   
\let\r=\rho         
\let\s=\sigma                   
\let\t=\tau
\let\w=\omega                                    \let\W=\Omega
\let\x=\xi                                       
\let\y=\upsilon                                  

 %
 %
\def\Box{\sqcap\llap{$\sqcup$}}
\def\lapp{\lower.4ex\hbox{\rlap{$\sim$}} \raise.4ex\hbox{$<$}}
\def\gapp{\lower.4ex\hbox{\rlap{$\sim$}} \raise.4ex\hbox{$>$}}
\def\con{\ifmmode\raise.1ex\hbox{\bf*}
          \else\raise.1ex\hbox{\bf*}\fi}
\def\bo{{\raise.15ex\hbox{\large$\Box\kern-.39em$}}}

\def\Imm{\mathop{\Im m}}

\def\dual{\relax\leavevmode\lower.9ex\hbox{\titlerms*}}
\def\define{\buildrel\rm def\over =}

\let\8=\otimes
 %
 %
 %
 %
\let\ba=\overline
\let\2=\underline

\let\Tw=\widetilde
 %
\def\like#1{\llap{\phantom{#1}}}
\def\dt#1{{\buildrel{\smash{\lower1pt\hbox{.}}}\over{#1}}}

\font\eightrm=cmr8
\def\6(#1){\relax\leavevmode\hbox{\eightrm(}#1\hbox{\eightrm)}}
\def\0#1{\relax\ifmmode\mathaccent"7017{#1}     
                \else\accent23#1\relax\fi}      
\def\7#1#2{{\mathop{\null#2}\limits^{#1}}}      
\def\5#1#2{{\mathop{\null#2}\limits_{#1}}}      
 %

 %

 %

 %

 %
\newbox\t@b@x
\def\rightarrowfill{$\m@th \mathord- \mkern-6mu
     \cleaders\hbox{$\mkern-2mu \mathord- \mkern-2mu$}\hfill
      \mkern-6mu \mathord\rightarrow$}
\def\tooo#1{\setbox\t@b@x=\hbox{$\scriptstyle#1$}%
             \mathrel{\mathop{\hbox to\wd\t@b@x{\rightarrowfill}}%
              \limits^{#1}}\,}
\def\leftarrowfill{$\m@th \mathord\leftarrow \mkern-6mu
     \cleaders\hbox{$\mkern-2mu \mathord- \mkern-2mu$}\hfill
      \mkern-6mu \mathord-$}
\def\froo#1{\setbox\t@b@x=\hbox{$\scriptstyle#1$}%
             \mathrel{\mathop{\hbox to\wd\t@b@x{\leftarrowfill}}%
              \limits^{#1}}\,}
 %
\def\frac#1#2{{#1\over#2}}
\def\frc#1#2{\relax\ifmmode{\textstyle{#1\over#2}} 
                    \else$#1\over#2$\fi}           
\def\inv#1{\frc{1}{#1}}                            
 %
\def\Claim#1#2#3{\bigskip\begingroup%
                  \xdef #1{\secsym\the\meqno}%
                   \writedef{#1\leftbracket#1}%
                    \global\advance\meqno by1\wrlabeL#1%
                     \noindent{\bf#2}\,#1{}\,:~\sl#3\vskip1mm\endgroup}

\def\QED{\rlx\hfill$\Box$\kern-7pt\raise3pt\hbox{$\surd$}\bigskip}
 %
 %

 %
\def\muthstrut{\vphantom1}
\def\mutrix#1{\null\,\vcenter{\normalbaselines\m@th
        \ialign{\hfil$##$\hfil&&~\hfil$##$\hfill\crcr
            \muthstrut\crcr\noalign{\kern-\baselineskip}
            #1\crcr\muthstrut\crcr\noalign{\kern-\baselineskip}}}\,}

 %
\def\YT#1#2{\vcenter{\hbox{\vbox{\baselineskip0pt\parskip=\medskipamount%
             \def\Box{$\sqcap\llap{$\sqcup$}$\kern-1.2pt}%
              \def\Z{\hfil\vskip-5.8pt}\lineskiplimit0pt\lineskip0pt%
               \setbox0=\hbox{#1}\hsize\wd0\parindent=0pt#2}\,}}}
\def\EU{\rlx\ifmmode \c_{{}_E} \else$\c_{{}_E}$\fi}
\def\TM{\rlx\ifmmode {\cal T_M} \else$\cal T_M$\fi}
\def\TW{\rlx\ifmmode {\cal T_W} \else$\cal T_W$\fi}
\def\CM{\rlx\ifmmode {\cal T\rlap{\bf*}\!\!_M}
             \else$\cal T\rlap{\bf*}\!\!_M$\fi}
\def\hm#1#2{\rlx\ifmmode H^{#1}({\cal M},{#2})
                 \else$H^{#1}({\cal M},{#2})$\fi}
\def\CP#1{\rlx\ifmmode\IP^{#1}\else\IP$^{#1}$\fi}
\def\cP#1{\rlx\ifmmode\IC{\rm P}^{#1}\else$\IC{\rm P}^{#1}$\fi}

\def\sll#1{\rlx\rlap{\,\raise1pt\hbox{/}}{#1}}
\def\Sll#1{\rlx\rlap{\,\kern.6pt\raise1pt\hbox{/}}{#1}\kern-.6pt}

\let\SSS=\scriptstyle

 %
 %

\def\CY{Calabi-\kern-.2em Yau}

\def\3{\ifmmode\ldots\else$\ldots$\fi}
\def\Z{\hfil\break\rlx\hbox{}\quad}
\def\3{\ifmmode\ldots\else$\ldots$\fi}
\def\?{d\kern-.3em\raise.64ex\hbox{-}}           
\def\9{\raise.43ex\hbox{-}\kern-.37em D}         
\def\ping{\nobreak\par\centerline{---$\circ$---}\goodbreak\bigskip}
 %
 %
\def\I#1{{\it ibid.\,}{\bf#1\,}}
\def\Pre#1{{\it #1\ University report}}

\def\NP#1{{\it Nucl.\,Phys.\,}{\bf#1\,}}
\def\PL#1{{\it Phys.\,Lett.\,}{\bf#1\,}}

\def\PRL#1{{\it Phys.\,Rev.\,Lett.\,}{\bf#1\,}}
\def\CMP#1{{\it Commun.\,Math.\,Phys.\,}{\bf#1\,}}
\def\CQG#1{{\it Class.\,Quant.\,Grav.\,}{\bf#1\,}}

 %
 %
 %
\baselineskip=13.0861pt plus2pt minus1pt
\parskip=\medskipamount
\let\ft=\foot
\noblackbox
\def\SaveTimber{\abovedisplayskip=1.5ex plus.3ex minus.5ex
                \belowdisplayskip=1.5ex plus.3ex minus.5ex
                \abovedisplayshortskip=.2ex plus.2ex minus.4ex
                \belowdisplayshortskip=1.5ex plus.2ex minus.4ex
                \baselineskip=12pt plus1pt minus.5pt
 \parskip=\smallskipamount
 \def\ft##1{\unskip\,\begingroup\footskip9pt plus1pt minus1pt\setbox%
             \strutbox=\hbox{\vrule height6pt depth4.5pt width0pt}%
              \global\advance\ftno by1\footnote{$^{\the\ftno)}$}{##1}%
               \endgroup}
 \def\listrefs{\footatend\vfill\immediate\closeout\rfile%
                \writestoppt\baselineskip=10pt%
                 \centerline{{\bf References}}%
                  \bigskip{\frenchspacing\parindent=20pt\escapechar=` %
                   \rightskip=0pt plus4em\spaceskip=.3333em%
                    \input \jobname.refs\vfill\eject}\nonfrenchspacing}}
 %
\def\Afour{\ifx\answ\bigans
            \hsize=16.5truecm\vsize=24.7truecm
             \else
              \hsize=24.7truecm\vsize=16.5truecm
               \fi}
\catcode`@=12
 %
 %
\def\rd{{\rm d}}

\def\db{\skew4\bar\partial}
\def\shB{{\cal B}^\sharp}
\def\shM{{\cal M}^\sharp}
\def\bm#1{\rlx\ifmmode\mathchoice
                      {\BM{#1}}
                      {\BM{#1}}
                      {\BM{\scriptstyle#1}}
                      {\BM{\scriptscriptstyle#1}}
                 \else{\BM{#1}}\fi}
\SaveTimber 
 %
 %
\font\subTitleFont=cmr10 scaled\magstep1
\Title{HUPAPP-93/1}
      {\vbox{\centerline{Spacetime Variable Superstring Vacua}
              \vglue3mm
             \centerline{\subTitleFont(Calabi-Yau Cosmic Yarn)}}}

 \centerline{\titlerms Paul S.~Green}                \vskip.2mm
 \centerline{Department of Mathematics}              \vskip 0mm
 \centerline{University of Maryland,
             College Park, MD 20742}                 \vskip 0mm
 \centerline{psg\,@\,lakisis.umd.edu}                \vskip 0mm
\vskip .2in
 \centerline{and}
\vskip .2in
 \centerline{\titlerms Tristan H\"ubsch\footnote{$^{\spadesuit}$}
      {\def\:{,\kern-.1em,\ignorespaces\nobreak}%
       On leave from the Institute ``Rudjer Bo\v skovi\'c'',
       Zagreb, Croatia.}}                              \vskip.2mm
 \centerline{Department of Physics}                    \vskip 0mm
 \centerline{Howard University, Washington, D.C.~20059}\vskip 0mm
\centerline{hubsch\,@\,scsla.howard.edu}
\vfill

\centerline{ABSTRACT}\vskip2mm
\vbox{\narrower\narrower\narrower\baselineskip=12pt\noindent
In a general superstring vacuum configuration, the `internal' space
(sector) varies in spacetime. When this variation is non-trivial only in
two space-like dimensions, the vacuum contains static cosmic strings
with finite energy per unit length and which is, up to interactions with
matter, an easily computed topological invariant. The total spacetime
is smooth although the `internal' space is singular at the center of
each cosmic string. In a similar analysis of the Wick-rotated Euclidean
model, these cosmic strings acquire expected self-interactions. Also, a
possibility emerges to define a global time in order to rotate back to
the Lorentzian case.}

\Date{April~'\number\yearltd \hfill}

\noblackbox
 %
 %
\nref\rSCS{B.R.~Greene, A.~Shapere, C.~Vafa and S.-T.~Yau:
      \NP{B337}(1990)1--36.}

\nref\rCHSW{P.~Candelas, G.~Horowitz, A.~Strominger and  E.Witten:
      \NP{B258}(1985)46--74.}

\nref\rDNAS{D.~Nemeschansky and A.~Sen: \PL{178B}(1986)365\semi
       P.~Candelas, M.D.~Freeman, C.~Pope, M.F.~Sohnius and
       K.S.~Stelle: \PL{177B}(1986)341.}

\nref\rLoopM{I.B.~Frenkel, H.~Garland and G.J.~Zuckerman: {\it
       Proc.\,Natl.\,Acad.\,Sci.\,USA\,\bf83}(1986)8442\semi
       M.J.~Bowick and S.G.~Rajeev: \PRL{58}(1987)535,
      \I{58(E)}(1987)1158, \NP{B293}(1987)348\semi
       L.~Alvarez-Gaum\'e, C.~Gomez and C.~Reina:
      \PL{190B}(1987)55\semi
       K.~Pilch and N.P.~Warner: \CQG{4}(1987)1183\semi
       P.~Oh and P.~Ramond: \PL{195B}(1987)130\semi
       D.~Harari, D.K.~Hong, P.~Ramond and V.G.J.~Rogers:
      \NP{B294}(1987)556.}

\nref\rBeast{T.~H\"ubsch: {\it \CY\ Manifolds---A Bestiary for
       Physicists} (World Scientific, Singapore, 1992).}

\nref\rMM{P.~Candelas, X.~de la Ossa, P.S.~Green and L.~Parkes:
      \PL{258B}(1991)118--126, \NP{B359}(1991)21--74.}

\nref\rRoll{P.~Candelas, P.S.~Green and T.~H\"ubsch: in ``Strings
       '88'', p.155, eds.~S.J.~Gates Jr., C.R.~Preitschopf and
       W.~Siegel (World Scientific, Singapore, 1989),
      \PRL{62}(1988)1956--1959, \NP{B330}(1990)49--102.}

\nref\rWeb{P.S.~Green and T.~H\"ubsch: \PRL{61}(1988)1163--1166,
      \CMP{119}(1989)431--441.}

\nref\rCecVaf{S.~Cecotti and C.~Vafa: Exact Results for Supersymmetric
       Sigma Models. \Pre{Harvard} HUTP-91/A062.}

\nref\rCYCI{T.~H\"ubsch: \CMP{108}(1987)291\semi
       P.~Green and T.~H\"ubsch: \CMP{109}(1987)99\semi
       P.~Candelas, A.M.~Dale, C.A.~L\"utken and R.~Schimmrigk:
      \NP{B298}(1988)493--525\semi
       B.~Greene, S.-S.~Roan and S.-T.~Yau: \CMP{142}(1991)245.}

\nref\rTYnc{G.~Tian and S.T.~Yau: Complete Ricci-Flat K\"ahler
       Manifolds.~I and~II. \Pre{Harvard} (1989).}

\nref\rABCG{M.~Atiyah, R.~Bott and L.~G\aa rding:
      {\it Acta Math.}~{\bf131}(1973)145\semi
       P.~Candelas: \NP{B298}(1988)458--492.}

\nref\rPeriods{P.~Berglund, P.~Candelas, X.~de la Ossa, A.~Font,
       T.~H\"ubsch, D.~Jan\v{c}i\'c and F.~Quevedo: Periods for
       Calabi-Yau and Landau-Ginzburg Models. {\it CERN Report}
       CERN-TH.~6865/93.}

\nref\rClem{H.~Clemens: {\it Publ.\,Math.\,IHES\ \bf58}(1983)19.}

\nref\rMan{S.~Mandelstam, \NP{B64}(1973)205.}
 %
 %
\newsec{Introduction, Results, and Summary}\noindent
The past several years have witnessed a demographic explosion of
superstring vacua with N=1 supergravity in 4-dimensional Minkowski
spacetime. To enable various anomaly cancellations, the model must
include a $c_L=c_R=9$ `internal' model, for which a geometric
description may not be known, but if it is---it must be a compact
complex Ricci-flat (\CY) 3-fold, possibly singular but in a rather
restricted way; see below. Most of these constructions are constant,
i.e., the `internal' sector of the theory is the same throughout the
4-dimensional spacetime.  Clearly, this is a rather special Ansatz and
attempts have been made to construct spacetime variable vacua
(see~\rSCS\ and the references therein); moreover, these may lead to
cosmologically more interesting and possibly realistic effective models.

In the most general situation, the internal space varies throughout
4-dimensional spacetime and couples to the geometry of the spacetime.

In this article, we will present some straightforward but general ways
of endowing families of \CY\ vacua with spacetime dependence. We
explore a class of superstring vacua in which the 10-dimensional
spacetime \BM{M} has the structure of a fibre space over the
4-dimensional spacetime $X$. That is, there exists a projection map
\eqn\eXXX{ \p ~:~ \BM{M} \longrightarrow X~, }
which associates a fibre ${\cal M}_x \subset \BM{M}$ to every point
$x\in X$. All fibres ${\cal M}_x = \p^{{-}1}(x)$ are compact complex
3-spaces, they vary over the 4-dimensional spacetime and we require the
total spacetime \BM{M} to be smooth. It turns out that in general
${\cal M}_x$ are smooth at every point $x\in X$ in the 4-dimensional
spacetime, except over a two-dimensional subset ${\cal S} \subset X$.

In such an Ansatz, we interpret $\cal S$ as the union of the world
sheets of one or more cosmic strings in the 4-dimensional spacetime
$X$. In the full 10-dimensional spacetime this looks as follows.
Consider the collection of fibres ${\cal M}_x$ over the cosmic string,
that is, for which $x\in{\cal S}\subset X$. Generically, each fibre has
one or perhaps several isolated singular points and a little thought
reveals that these singular points sweep out a 2-dimensional surface
$\Tw{\cal S} \subset {\bm M}$, of which ${\cal S} = \p(\Tw{\cal S})$
is the projection to the 4-dimensional spacetime $X$; see Fig.~1.


If the total 10-dimensional spacetime \BM{M} is smooth and has a
Ricci-flat metric, ${\bf g}$, it qualifies as a superstring
vacuum~\rSCS\ft{This is easy to show along the arguments of
\refs{\rCHSW,\rDNAS}; in fact, Ricci-flatness appears to be a general
requirement for string theories~\rLoopM.}. At every point $m\in \BM{M}$
where the projection $\p$ is non-singular and using the metric ${\bf g}$,
we can define the `horizontal' tangent space to be the orthogonal
complement of the tangent space to the `vertical' fibre ${\cal M}_x$.
Furthermore, each (compact, \CY) fibre ${\cal M}_x$ for which $x$ is not a
singular value of $\pi$ admits a holomorphic 3-form, and therefore a
Ricci-flat metric. However the restriction of ${\bf g}$ to ${\cal M}_x$
is not in general Ricci flat. Finally, we can define a 4-dimensional
space-time metric by restricting the 10-dimensional metric to the
horizontal subspaces and integrating over the fibers.

It turns out that vacua of this kind can be constructed in abundance
using certain techniques of (complex) algebraic geometry. In particular,
following the analysis of Ref.~\rSCS, we first focus on the case where
the `internal' \CY\ space varies only in two space-like dimensions. The
`internal' space then becomes singular at a prescribed number of points
of this surface, each of which sweeps out a {\it static} cosmic string
in the third spatial dimension. Certain physically relevant properties
of these cosmic strings can be determined straightforwardly.

The presence of these strings curves the spacetime\ft{Unless otherwise
specified, by `spacetime' we mean the 4-dimensional one.} and induces a
spacetime metric, which in fact turns out to be related to a flat
metric by a conformal factor whose logarithm is the K\"ahler potential
of the Weil-Petersson-Zamolodchikov metric. Several local properties of
this spacetime metric are then computed easily. In the general case,
when the `internal' space varies over the entire 4-dimensional
spacetime, we have not found any comparably simple characterization of
the 4-dimensional space-time metric.

We proceed as follows~: Section~2 presents the problem from the point
of view of a ``4-dimensional physicist'', analyzing the effect of
varying the `internal' space over spacetime as a 4-dimensional field
theory problem. Section~3 offers a more general point of view by
considering the geometrical structure of the total 10-dimensional
spacetime. Some simple calculations are described in Section~4 and
Section~5 presents some concrete models and corresponding calculations.
Section~6 remarks on certain unusual global properties of such
configurations and a preliminary analysis of the fully variable
10-dimensional spacetime case and our closing remarks are presented in
Section~7.

\newsec{Varying Vacua in Spacetime}\noindent
While spacetime variable superstring vacua are clearly more general
solutions of the superstring theory, such a variation may produce
physically unacceptable effects: physical observables such as the mass
of the electron may vary too much. Nevertheless, we consider varying
the `internal' sector over spacetime and leave checking the physical
acceptability {\it a posteriori}.

\subsec{Variable `internal' space field theory}\noindent
Suitable `internal' models are 2-dimensional $(2,2)$-superconformal
$c_L=c_R=9$ field theories, all of which have marginal operators.
Through these, any such model may be deformed into a neighboring one,
that is, such operators chart a local deformation space. In spacetime
variable superstring vacua, these marginal deformations acquire
spacetime dependence.

Once the parameters of the `internal' model have become spacetime
variable, so-called moduli fields $t^\a(x)$, they are
naturally interpreted as maps that immerse the spacetime into the
moduli space $\cal B$ of the `internal' model. Their dynamics is then
governed by the usual $\s$-model (harmonic map) action. To discuss
this, focus on the natural interaction of $t^\a$ with gravity which
follows simply by requiring the action to be invariant under general
coordinate reparametrizations. So consider
\eqn\eEffAct{ {\cal A}_{\rm eff.}~~ = ~~
       \int \rd^4x~ \sqrt{-g}\> \Big[\> - \inv2 R~ -
      ~G_{\a\bar\b}(t,\bar t)\> g^{\m\n} \> \vd_\m t^\a\,
                            \vd_\n t^{\bar\b} \>\Big]~, }
using the facts that the moduli space of the `internal' models is
complex and that the natural (Weil-Petersson-Zamolodchikov) metric $G$
on the moduli space is Hermitian. $g^{\m\n}$ is the spacetime inverse
metric, $g\define \det[g_{\m\n}]$ and $R$ is the spacetime scalar
curvature.  As usual, the equations of motion for $t^\a$ are
\eqn\eEqMotn{ g^{\m\n}\big[\> \nabla_\m \nabla_\n t^\a~
    + ~{\mit\G}_{\b\g}^\a(t,\bar t)\> \vd_\m t^\b\,\vd_\n t^\g
                                               \>\big]~~ = ~0~, }
where $\nabla_\m$ is the spacetime gravitationally covariant derivative
and ${\mit\G}_{\b\g}^\a= \inv2 G^{\a\bar\d}\vd_{(\b\like{$\SSS\bar\d$}}
G_{\g)\bar\d}$ is the Cristoffel connection on the moduli space derived
from the Weil-Petersson-Zamolodchikov metric. Generally, solutions of
Eq.~\eEqMotn\ are minimal submanifolds of moduli space, i.e.,
spacetimes embedded in the moduli space with least action. Given such a
solution, we then need to solve the Einstein equations
\eqn\eEqEins{ R_{\m\n} - \inv2 g_{\m\n}\> R~~
                  = ~~T_{\m\n}[\,t,\bar t\,]~, }
for the spacetime metric $g_{\m\n}$. Here, as usual,
\eqn\eEMTens{  T_{\m\n}[\,t,\bar t\,]~~ =  ~~-G_{\a\bar\b}
     \big[\, \vd_{(\m}\,t^\a\> \vd_{\n)}\,t^{\bar\b}
     -\inv2 g_{\m\n}(g^{\r\s}\vd_\r t^\a\,\vd_\s t^{\bar\b}) \,\big] }
is the energy-momentum tensor of the moduli fields $t^\a(x)$, which
are being treated as matter from the point of view of gravity even
though they describe the vacuum. In fact, because of the
gravitationally covariant derivatives $\nabla_\m$ in~\eEqMotn, which
involve a choice of a connection over the full 4-dimensional
spacetime, Eqs.~\eEqMotn--\eEMTens\ form a nontrivially coupled system.
As usual, we would treat them perturbatively, solving first
Eq.~\eEqMotn\ in the flat background ($\nabla_\m=\vd_\m$) to determine
the vacuum, then solving Eq.~\eEqEins\ for the metric and finally
verifying that the feedback of Eq.~\eEqEins\ into Eq.~\eEqMotn\ is
negligible, or as in our main case---zero. \ping

For a simple but rather abundant class, assume that $t^\a$ is static
and varies only in two space-like dimensions, which for each fixed
$x_3$ span a space-like infinite surface $\cal Z$. This simplifies the
problem, as seen by combining these two space-like coordinates into a
complex coordinate, whereupon the action~\eEffAct\ and Eq.~\eEqMotn\
become %
\eqna\eEActz
$$ \eqalignno{
 {\cal A}_{\rm eff.} &=
   ~~ -\inv2\int \rd^4x~ \sqrt{-g}\> R~~ -
      ~~\int\rd t\,\rd x_3\, {\cal E} ~,              & \eEActz{a}\cr
 {\cal E}~~          &\define
   ~~ \int_{\cal Z} \rd^2z~\sqrt{-g}\, G_{\a\bar\b}\, g^{z\bar z}\,
       \big(\,\vd_z t^\a\,\vd_{\bar z}t^{\bar\b} +
        \vd_{\bar z}t^\a\,\vd_z t^{\bar\b} \,\big)~, & \eEActz{b}\cr}
$$
and, respectively (neglecting gravity),
\eqn\eEqMoFl{ \vd_z \vd_{\bar z} t^\a~
    + ~\vd_z t^\b\,{\mit\G}_{\b\g}^\a\,\vd_{\bar z}t^\g ~~ = ~0~. }
Indeed, this equation of motion for the moduli fields is most easily
solved by taking the $t^\a$ to be {\it arbitrary} (anti)holomorphic
functions of $z$. We thus consider holomorphic maps of a complex plane
into the moduli space of the `internal' $c_L=c_R=9$ model. By adding a
point at spatial infinity, we compactify the surface $\cal Z$ into a
2-sphere, $S^2 = \CP1$, and will hence consider holomorphic mappings
${\cal Z}^c=\CP1 \to \bar{\cal B}$, where $\bar{\cal B}$ is a suitable
(partial) compactification of the moduli space $\cal B$.

For holomorphic $t^\a$, the expression~\eEActz{b} simplifies in an
important way: the second term drops out and $\cal E$ is the integral
over ${\cal Z}^c$ of the (1,1)-form associated to the pull-back metric
$[t^*(G)]_{z\bar z} = [\vd_z t^\a G_{\a\bar\b}\vd_{\bar z}t^{\bar\b}]$.
Since the Weil-Petersson-Zamolodchikov metric $G_{\a\bar\b}$, is
actually K\"ahler,
\eqn\eEnDens{ {\cal E}_{z\bar z}~~ \define
   ~~\vd_z t^\a\,\vd_{\bar z}t^{\bar\b}\> G_{\a\bar\b}~
  = ~\vd_z\, \vd_{\bar z}\, K~, }
and so
\eqn\eEisC{ {\cal E}~~ = ~~i\int_{{\cal Z}^c}\vd\db\>K~~
                         \define ~2\p \deg t({\cal Z}^c)~. }
Looking back at Eq.~\eEqMoFl, we see that the trivial solutions are
$t=\hbox{\it const.}$, for which ${\cal E}=0$. In less trivial cases,
we see that ${\cal E}\ne0$ is the total energy per unit length (along
$x_3$) contained in the fields $t^\a$.
As long as the singularities of the integrand are not too bad,
\eEisC is invariant against holomorphic deformations and $\deg t$
is an integer. In most of the cases we will consider, $t$ factors
through a projective space (e.g., the space of quintics in
$\CP4$), and $\deg t$ is the degree of $t$ as a polynomial mapping.

Having tentatively compactified $\cal Z$ into a \CP1, $t({\cal Z}^c)$
represents a rational curve (or its multiple cover) in some partial
compactification of the moduli space; the constant (trivial) solutions
corresponding to mapping the whole ${\cal Z}^c$ into a single point of
the moduli space. Eventually, one point on this (complex) curve will be
identified with the infinity and removed, restoring our open space-like
surface $\cal Z$ and we will later obtain some sufficient conditions for
this to produce a consistent superstring solution.

\subsec{Spacetime structure in the vacuum}\noindent
The moduli space of $c_L=c_R=9$ $(2,2)$-superconformal models is
spanned by two types of variations, corresponding to deformations of
the complex structure and (complexified) variations of the K\"ahler
class\ft{We adhere to the \CY\ geometrical vernacular although not every
$c_L = c_R = 9$ (2,2)-supersymmetric model has a known geometrical
interpretation. Our analysis easily extends to such models also.}. The
mirror-map equates the physically natural metric structure, so-called
`special geometry', of the two sectors in this moduli space. It
suffices therefore to consider only one of them and we choose to
discuss deformations of the complex structure, these being simpler to
describe exactly. The anticipated rigorous establishment of mirror
symmetry will then ensure that the analogous description  holds for
variations of the (complexified) K\"ahler class.

For a deformation class of \CY\ 3-folds
 $\{{\cal M}_t\,,~t\in{\cal B}\}$, the space $\cal B$ is complex and
moreover K\"ahler. Locally at $t$, it is parametrized by
 $H^1({\cal M}_t,{\cal T}_{{\cal M}_t})$, which in turn equals
 $H^{2,1}({\cal M}_t)$ since $\cal M$ is \CY. That is,
 $H^1({\cal M}_t,{\cal T}_{{\cal M}_t})$ is tangent to $\cal B$ at $t$.
There are no non-constant holomorphic maps from ${\cal Z}^c$ to $\cal B$.
However if we enlarge $\cal B$ to $\bar{\cal B}$, the completion of
 $\cal B$ in the Weil-Petersson-Zamolodchikov metric, the situation
changes.  As we shall see below, there are many cases of interesting
holomorphic maps from ${\cal Z}^c$ (in some cases also from $\cal Z$)
to $\bar{\cal B}$. Because $\shB\define{\bar{\cal B}}{-}{\cal B}$ has
codimension one in $\bar{\cal B}$, the set of points mapping to $\shB$
must be finite when the domain is ${\cal Z}^c$ and is so in the cases
we consider even when the domain is the noncompact $\cal Z$. $\shB$ has
a subset $\shB_{\rm con}$ whose points correspond to conifolds,
varieties which are smooth except for a single node. $\shB_{\rm con}$
is open in $\shB$ and can be shown in most cases to be dense in $\shB$.
When this is the case, it can be arranged that all the exceptional
points map to
 $\shB_{\rm con}$, and in fact this will then be the generic situation.

For sub-generic\ft{Our use of the term
``generic'' in the sequel will refer to the situation in which there
exist maps from ${\cal Z}^c$ to ${\cal B}\cup\shB_{\rm con}$, although
much of the analysis carries over into  cases where worse
singularities than nodes, or even clusters of nodes, are unavoidable.}
choices of the holomorphic $t$, the `internal' \CY\ 3-fold will acquire
more and/or worse singularities. For example, if the image $t({\cal
Z}^c)$ is arranged to intersect $\shB$ at an $n$-fold normal crossing
$t^\sharp$, the {\it conifold} $\shM_{t^\sharp}$ will have $n$
nodes\ft{However, requiring additional symmetries of the deformed
potential may be required, e.g., for construction of the mirror
model~\rMM\ and may very well enforce the occurrence of singularities
worse than nodes.}.

These special points in $\cal Z$, where the `internal' space
singularizes, sweep a filamentary structure along the third
space-like coordinate, identified as cosmic strings. Note that these
substantially differ from the cosmic strings in Grand-unified
theories: there $\cal E$ is divergent, although typically only
logarithmically. Here, $\cal E$ is finite and moreover a topological
invariant~\eEisC.
\ping

Ref.~\rSCS\ analyzed in great detail the toy model in which the \CY\
3-fold is chosen to be $T \times\rm K3$ (or $T\times T\times T$) and
only the complex structure of the torus $T$ is allowed to vary.
There, $T$ can have only nodes. Also, we have that $G_{\t\bar\t} =
\inv4(\Imm\t)^{{-}2} = -(\t-\ba\t)^{{-}2}$, so
\eqn\eEnDensT{ {\cal E}_{\rm torus}~~ = ~~{i\over2} \int_{\cal Z}
                          {\rd\t\,\rd\ba\t \over (\t-\ba\t)^2}~~
                      = ~~{ m\over 24 }\cdot4\p~, }
where $m$ is the multiplicity of the mapping $\t:{\cal Z}\to F$ of the
compactified space-like surface $\cal Z$ to the fundamental region $F$
of the torus moduli space.  Suffice it here to note that for
consistency reasons, $m$ is a multiple of 6~\rSCS.

For the general case, where the variable `internal' space $\cal M$ is
an irreducible \CY\ 3-fold, the Weil-Petersson-Zamolodchikov
metric equals
\eqn\eWPZmet{ G_{\a\bar\b}(t,\bar t)~~ \define
             ~~\vd_\a\,\vd_{\bar\b}\>K~,\qquad
           K~~ \define
             ~~-\log\big(i\!\int_{{\cal M}_t} \kern-.8em
                 \W_t\wedge\ba{\W_t}\,\big)~, }
where $\W_t$ is the holomorphic 3-form on ${\cal M}_t$ at $t\in\cal B$.
Consider the generic case when $\cal M$ develops one or perhaps a few
nodes at some special points $z_i\in\cal Z$. We then have that locally
around each special point $t_i = t(z_i^\sharp)$ where ${\cal M}_t$
becomes singular,
\eqn\eAsymp{ e^{-K}~~ \sim
          ~~a(\t_i,\bar\t_i) + b(\t_i,\bar\t_i)|\t_i|^2\log|\t_i|~,
             \qquad \t_i \define t-t_i~,}
and where we know that $a(\t_i,\bar\t_i)$ and $b(\t_i,\bar\t_i)$ are
bounded, non-zero and smooth $C^{\infty}$ functions around $\t_i=0$.
For example, in a concrete model of Ref.~\rRoll, $a(\t,\bar\t) =
a_0(1-|\t|^2)$ and $b(\t,\bar\t)=b_0$ to leading order in $\t_i$ and
$a_0$ and $b_0$ are readily calculable constants. This implies that
$G_{\t_i\bar\t_i} \sim \log|\t_i|^2$. So, in the total energy per unit
length
\eqn\eXXX{ {\cal E}~~ =
         ~~-i\!\int_{\cal Z} \vd\,\db\>
              \log\big(i\!\int_{{\cal M}_{t(z)}} \kern-.8em
                         \W_{t(z)}\wedge\ba{\W_{t(z)}}\,\big)~, }
the integrand
\eqn\eXXX{ {\cal E}_{z\bar z}~~ = ~~\vd_z\vd_{\bar z}K~~ =
           ~~-\vd_z\vd_{\bar z} \log\big(i\!\int_{{\cal M}_{t(z)}}
              \kern-.8em \W_{t(z)}\wedge\ba{\W_{t(z)}}\,\big) }
diverges logarithmically at the locus of each string.  So far, the
behavior of $G_{\a\bar\b}(t,\bar t)$ was in general not known away from
$t_i$, the value(s) of moduli corresponding to conifold
singularization(s). However, the techniques of Ref.~\rMM\ lend
themselves for straightforward generalization and a powerful and rather
general machinery is currently being developed to this end~\rPeriods.

So far, $G_{\a\bar\b}$ has been computed only for one one-parameter
family of models, in Ref.~\rMM. Fig.~4.1 there plots the
Weil-Petersson-Zamolodchikov metric against the only parameter $\j$.
{}From this plot, it is clear that the regions $|\j|>1$ and $|\j|<1$
provide commensurate contributions to $\cal E$ ($\int_{|\j|>1} \vd\db K
\approx 4\int_{|\j|<1} \vd\db K$), and that the small region
$|1-\j|\lapp\inv{16}$ contributes a similar (convergent) amount owing to
the logarithmic divergence at $\j=1$ where the \CY\ 3-fold develops 125
nodes. So, the energy per unit length of the field $t(z)$ is  stored
partly away from the string and partly around it. This contradicts the
na\"\ii ve expectation that the energy density of a string ought to be
concentrated around its locus. As we shall see below, this is due in
part to the fact that there is a highly charged string ``hiding'' at
infinity, and in part to the fact that we are not yet comparing the
curvature to the physical space-time metric.

In a generic situation, there are $m$ points $z_i^\sharp\in\cal Z$ where
$\shM_{t(z_i)}$ is a conifold with a single node and worse
singularizations of $\cal M$ do not occur. Assuming that the energy per
unit length is distributed similarly to the case of Ref.~\rMM, we expect
that there would be a more or less uniform `background' energy density
with logarithmically divergent spikes at the loci of the cosmic strings.
Now, if the point at infinity corresponds to a smooth rather than a
singular \CY\ 3-fold, then the background energy density will fall off
exponentially with the distance. This follows easily from two
observations pertaining to large $|z|$, i.e., far away from the strings.
The first is that the energy density falls of in this case as ${1\over
z\bar z}$; the second is that, as we shall see below, the physical
metric is asymptotically proportional to  $dz\,d\bar z\over z\bar z$.
It follows that the  physical  distance is asymptotically proportional
to $\log|z|$, so that the energy density falls of exponentially with
the physical distance, as asserted.

{}From Eq.~\eEisC, the {\it total} energy per unit length is a
deformation invariant, $\inv2 C_1{\cdot}4\pi$, where $C_1$ is the
degree of the mapping from our projective line to the projective
space parametrizing the complex structures. For example, in the
torus case, the $m=12$-string case corresponds to a $C_1=1$ (linear)
mapping and can easily be described by embedding the torus as a cubic
curve in \CP2
\eqn\eTorLin{ Q_0(u,v,w) + z\, Q_1(u,v,w)~ = ~0~,\qquad
                 (u,v,w) \in \CP2~, }
where $z\in{\cal Z}=\CP1$ and $Q_0$ and $Q_1$ are homogeneous cubics.

\subsec{The spacetime metric}\noindent
Quite naturally, one assumes the spacetime metric to be specified as
\eqn\eXXX{ \rd s^2~~ =
  ~~-\rd t^2~ + ~e^{\f(z,\bar z)}\rd z\rd\bar z~ + \rd x_3{}^2~. }
This Ansatz has the virtue of leaving the expression~\eEActz{b}
unchanged, since the relation $\sqrt{-g}\,g^{z\bar z}=1$ obtains. So
does therefore also Eq.~\eEqMoFl, and $t^\a(z)$ can again be chosen to
be arbitrary holomorphic functions. With this, the energy-momentum
tensor and the Einstein equations~\eEqEins\ simplify, the latter of
which reduce to a single independent equation
\eqn\eXXX{
   \Big[\, R_{z\bar z}-\inv2g_{z\bar z}R =
           \inv2\vd_z \vd_{\bar z}\f \,\Big]~~
 = ~~\Big[\, T_{z\bar z} = -\inv2\vd_z \vd_{\bar z}K \,\Big] ~. }
This suggests the simple solution $\f=-K$, that is, with~\eWPZmet,
\eqn\eSpTiMe{ \rd s^2~~ = ~~-\rd t^2~ +
    ~\big(i\!\int_{{\cal M}_t} \kern-.8em
      \W_t\wedge\ba{\W_t}\,\big)\rd z\rd\bar z~
     + \rd x_3{}^2~. }

Now, as the logarithm of the K\"ahler potential for the
Weil-Petersson-Zamolodchikov metric, $(i\int\W_t\,\ba{\W_t})$ is
defined only up to the modulus of a holomorphic function in $t$. In
other words, there is a `gauge transformation', generated by the
freedom to rescale $\W_t \to f(t)\W_t$ by any non-zero holomorphic
function $f(t)$. As a component of the spacetime metric, however, this
ambiguity is rather undesired. In the torus case, as discussed in
Ref.~\rSCS, the remedy was found by multiplying $e^{-K}=\Imm\t$ by
$\h^2\ba{\h}^2$, where $\h$ is the Dedekind modular function. Since we
do not yet have a detailed Teichm\"uller theory of \CY\ spaces, we must
proceed differently. Surprisingly perhaps, the ambiguity will be
resolved up to a constant factor by requiring  that the integral in
\eSpTiMe\ be finite and non-vanishing except at the point corresponding
to spatial infinity, as we shall see in more detail in the next section.

Finally, we note that, unlike the torus case where $e^{{-}K}$ blows up
at the cosmic strings~\rSCS, here $g_{z\bar z} = e^{-K}$ has a finite
value and a continuous first derivative at any $t_i$.
It is straightforward that the distance function $\int_0^r
\rd|z|\sqrt{g_{z\bar z}}$, with the results in Eqs.~\eSpTiMe, \eWPZmet\
and \eAsymp, is nearly linear for small $r$. This property is in sharp
contrast with the torus case and is at the root of most novel features
of the irreducible\ft{This `irreducibility' means that the holonomy of
the `internal' space is $SU(3)$ rather then a subgroup thereof; this is
also the condition for spacetime N=1 rather then extended
supersymmetry.} \CY\ 3-fold moduli space.

\subsec{Curvature and Yukawa couplings}\noindent
Since the spacetime metric behaves rather nicely at the locus of a
cosmic string, one may doubt the validity of the identification of this
filamentary structure with cosmic strings. Recall, however, that the
energy density does diverge at the loci $t_i$ of the
strings, although only logarithmically.

In fact, there are two important identities. On one hand, practically
by definition, the energy distribution in~\eEnDens, ${\cal E}_{z\bar z}
\define {-}\vd_z\vd_{\bar z} \log\big(i\int \W\wedge\ba\W\big)$, equals
the Weil-Petersson-Zamolodchikov metric. Therefore, it is everywhere
positive. On the other hand, ${\cal E}_{z\bar z}$ is precisely the
(Ricci) curvature for our spacetime metric~\eSpTiMe. The spacetime
scalar curvature is then
 $g^{a\bar a}{\cal E}_{a\bar a} = e^K{\cal E}_{z\bar z}$
and diverges logarithmically at the locus of these cosmic strings.

As easily seen from Eq.~\eSpTiMe, the spacetime (Ricci) curvature
equals the Weil-Petersson-Zamolodchikov area and so is positive over
the space-like surface $\cal Z$. Note also that the extrinsic curvature
of this surface in the 4-dimensional spacetime vanishes; transversally
to $\cal Z$, spacetime is flat.

Besides this metric structure, there is another quantity which (very
effectively) detects the locus of the cosmic strings---the Yukawa
couplings. One of the $b_{2,1}$ {\bf27}-fields of the effective model,
$\J_i$, will correspond to the deformation of the complex structure
proportional to $\t_i=(t-t_i)$. As in Ref.~\rRoll, one may compute that
the (unnormalized, but otherwise exact) Yukawa coupling behaves like
$\k_{iii} \sim (C/\t_i)$, for some constant $C$. It can also be shown
that, to leading order, the Yukawa coupling of $\J_i$ with any
other {\bf27}-field vanishes. The kinetic term of $\J_i$ is the same as
for $t^\a$ in~\eEActz{b} and so the leading term is $\log|\t_i|$.
Redefining $\J_i \to \J_i(\log|\t_i|)^{-1/2}$, the kinetic term becomes
the canonical one. The Yukawa coupling then becomes
$\big(C/[\t_i(\log|\t_i|)^{3/2}]\big)$ and still diverges: $\J_i$ is
(infinitely) strongly coupled at $\t_i=0$. Alternatively, rescale $\J_i
\to \t_i{}^{1/3}\J_i$, so that the Yukawa coupling becomes constant.
Then, however, the kinetic term becomes $|\t_i|^{2/3}\log|\t_i|$ and
dies out as $\t_i\to0$, turning $\J_i$ into a non-propagating field.

Either way, the dynamics of the $\J_i$ detects the cosmic string rather
drastically\ft{It has been suggested that the singularizations of the
`internal' space such as at those that occur at the locus of these
cosmic strings may interface to `internal' spaces of distinct
topology~\refs{\rRoll, \rWeb}. More recently, this possibility finds
support in some exact results~\rCecVaf.}---it becomes ``frozen''.
Again, this is rather different than in the cosmic strings of
Grand-unified theories, where some (Higgs) scalars become {\it
massless} at the core of the string. Finally, the above conclusions
rely on the exceptional features of `special geometry' to protect the
general relation between Yukawa couplings and the
Weil-Petersson-Zamolodchikov metric.
However, the spacetime singularity at the cosmic strings indicates that
there the vacuum configuration no longer makes sense as a
compactification, limiting the validity of the above analysis.
We thus seek a more general description.

\newsec{Advantages of a Bigger Vantage Point}\noindent
Reflect for the moment on the vacua described above. In essence, the
10-dimensional spacetime \BM{M} has become the product of a
2-dimensional Minkowski spacetime (spanned by $x_3$ and time) and an
8-dimensional compact space $M^c$. This 8-dimensional space has the
structure of a fibre space, where the compactified space-like surface,
${\cal Z}^c=\IP^1$, serves as a base and the \CY\ 3-folds ${\cal M}_z$
are associated to each $z\in{\cal Z}^c$ in such a way that the complex
structure of ${\cal M}_z$ varies holomorphically with $z$. We now turn
to some general methods of constructing suitable K\"ahler complex
4-dimensional spaces $M^c$, from which then non-compact \CY\ 4-folds
can be manufactured with ease.

\subsec{Constructing spacetime variable vacua}\noindent
Start with a non-singular complex projective four-fold \BM{X} the
anti-canonical bundle $K_{\bm X}\!\con$ of which is non-trivial and is
generated by global holomorphic sections. That is, if $\f(x)$ is such a
section (`polynomial') of $K_{\bm X}\!\con$, the zero-set in \BM{X} of
$\f(x)=0$ is a \CY\ 3-fold. Numerous simple examples of such spaces are
provided in the collection of \CY\ 3-folds constructed as complete
intersections of hypersurfaces in products of (weighted) complex
projective spaces~\refs{\rBeast, \rCYCI}. Each of these
may be thought of (typically in several different ways) as a simple
hypersurface in some 4-fold \BM{X}, obtained as the intersection of all
but one hypersurface.

The above conditions on $K_{\bm X}\!\con$ ensure that it has at least
two independent global holomorphic sections; let $\f_1,\f_2$ be two
such sections. We then form a 2-dimensional linear space $V = \{
z_1\f_1 + z_2\f_2 \}$. Each element from this space defines a \CY\
3-fold  as the zero-set, which remains the same if we rescale the
section by an arbitrary complex number. The parameters $z_1,z_2$ may
therefore be considered as projective coordinates on $P(V)=\IP^1_z$,
readily identified with the compactified space-like surface ${\cal
Z}^c$.

In other words, we consider $M^c$, the space of solutions of
\eqn\eDefEq{ \F(x,z)~ \define
             ~z_1\,\f_1(x)~ + ~z_2\,\f_2(x)~ = ~0~,\qquad
              x\in {\bm X}~,~~ z\in\CP1={\cal Z}^c~, }
which is simply a 4-fold in $\BM{X} \times \IP^1$. If $\p$ denotes the
projection on ${\cal Z}^c=\IP^1$, the inverse image $\p^{{-}1}(z)$ is
the subspace in $M^c$ which is projected to $z$ and is easily seen to
be the \CY\ 3-fold ${\cal M}_z \subset\BM{X}$. This provides the 4-fold
$M^c$ with the structure of a fibre space with \CY\ 3-folds
${\cal M}_z$ fibred over ${\cal Z}^c$.

In general, most choices of $\f_1,\f_2$ will yield a non-singular $M^c$.
However, there will always be a finite set of singular fibres $\shM_i$
at some $z^\sharp_i$, the number of which (counting with appropriate
multiplicities) can be predicted from  the choice of \BM{X}, using
elementary techniques of algebraic geometry.
\ping

The resulting 4-fold $M^c \subset \BM{X} \times \CP1$ itself is not \CY.
Its first Chern class equals one half of the first Chern class of
${\cal Z}^c = \CP1$. That is, the canonical bundle of $M^c$, $K_{M^c}$,
is isomorphic\ft{For $\F(x,z)=0$ to define a \CY\ 4-fold, $\F(x,z)$
 would have to be quadratic in $z$, since $c_1(\IP^1)=2J_{\IP^1}$; our
$\F(x,z)$ is only linear, so that $c_1(M^c)=J_{\IP^1}$.} to
$\pi^*\big({\cal O}_{\IP^1}(-1)\big)$, where $\p:M^c \to \CP1$
denotes the projection on the $z$-coordinates. Let
\eqn\eResIso{ \vp ~:~ \p^*\big({\cal O}_{\IP^1}(-1)\big)~
                    \buildrel{\sim}\over\longrightarrow
                    ~K_{M^c} }
denote this isomorphism (soon to be made more explicit). Recall that
$K_{M^c} \define \det{\cal T}_{M^c}\!\con = \W^4_{M^c}$ is the bundle
of holomorphic 4-forms over $M^c$.

{}From this compact 4-fold $M^c$, we will obtain a non-compact 4-fold $M$
by excising a fibre ${\cal M}_{z_\infty}\subset\BM{X}$ and the
corresponding point $z_\infty\in{\cal Z}^c$, identified as the spatial
infinity, and show that $M$ is a non-compact \CY\ 4-fold by
constructing explicitly a nowhere zero holomorphic 4-form.\bigskip

Let $z_1,z_2$ be homogeneous coordinates on ${\cal Z}^c = \CP1$, chosen
so that the point $z_\infty = (1,0)$ corresponds to the spatial
infinity. Then there is a unique meromorphic section of
${\cal O}_{\IP^1}(-1)$ with a pole precisely at $z_\infty$, which may
be written as $\m={c\over z_2}$, where $c$ is an irrelevant
constant. Note that $\m$ is non-zero over ${\cal Z}^c$ and blows up
only at spacetime infinity. Then, $\vp(\m)=\BM{\mit\W}$ is a nowhere
zero holomorphic 4-form over $M \define M^c {-} \p^{-1}(z_\infty)$.
Thus, as promised, the non-compact space $M$ admits a nowhere-zero (and
finite) holomorphic 4-form, which ensures that there is a Ricci-flat
K\"ahler metric on it~\rTYnc, provided $z_\infty$ is a non-singular
value of $\pi$.

As the above argument may appear too abstract, we also construct a more
explicit representative for $\BM{\mit\W}$, using a by now well known
residue formula~\rABCG. On $\IP^4 \times \IP^1$, there is a holomorphic
$5$-form,
    $\w \define \inv{5!}\e_{ijklm}x^i\rd x^j\rd x^k\rd x^l\rd x^m\,\*\,
              \inv{2!}\e_{ab}z^a\rd z^b$.
The quantity $\w / \F(x,z)$ has a simple pole at $M^c \subset \IP^4
\times \IP^1$, where $\F(x,z)=0$. This then serves as an integral
kernel to define forms on the compact submanifold $M^c$. That is, for
suitable $f(x,z)$, one calculates the residue of $f(x,z)\,\w/\F(x,z)$
at the locus of $\F(x,z)=0$, $M^c$. Now, note that $\w/\F(x,z)$ is
homogeneous of degree 0 over $\IP^4$, but of degree $+1$ over $\IP^1$.
To make it into an invariant form, $f(x,z)$ must be chosen homogeneous
of degree $(0,-1)$. Indeed, the meromorphic section $\m$ exactly suits
this purpose. If point by point $(x,z)\in M^c$, $\G$ is a circuit in
$\IP^4 \times \IP^1$ about $(x,z)$, the quantity
\eqn\eResOm{ {\eqalign{ \BM{\mit\W}(x,z)~
 &\define ~~{\rm Res}_{(x,z)}
   \Big[\,{c\over z_2} {\w\over\F(x,z)} \,\Big]~,                 \cr
 &= ~~\lim_{\G\to(x,z)}\oint_{\G} {c\over z_2} {\w\over\F(x,z)}~,
                                            \qquad (x,z)\in M^c~, \cr}}}
is easily seen to be a nowhere zero holomorphic 4-form, which blows up
precisely over $z_\infty$.

This residue calculation explicitly realizes the isomorphism~\eResIso:
to every (meromorphic) section of ${\cal O}_{\IP^1}(-1)$ over
${\cal Z}^c$, it assigns a meromorphic 4-form over $M^c$, that is
section of $K_{M^c}$. The only pole of \BM{\mit\W} is precisely at
spatial infinity, $z_\infty\in{\cal Z}^c$. Thus, over
\eqn\eXXX{ M~~ \define ~M^c~ - ~{\cal M}_\infty~;\qquad
          {\cal M}_\infty=\pi^{-1}(z_\infty)}
\BM{\mit\W} is nowhere-zero and finite---as promised.
\bigskip

On the other hand, $\m^2=c^2(z_2)^{{-}2}$ lives in
${\cal O}_{\IP^1}(-2) = K_{\IP^1} = \W^1_{\IP^1}$ and may therefore be
identified with $\rd z$ over ${\cal Z} \define {\cal Z}^c {-}
z_\infty$. Then for each non-singular fibre ${\cal M}_z = \p^{-1}(z)$,
$z\ne z_\infty$, there is a holomorphic 3-form $\W_z$, defined by
\eqn\eFactOm{ \m^2\wedge\W_z = \rd z \wedge \W_z~
                          = ~\vp(\m) = \BM{\mit\W}~. }
This clearly defines the nowhere-zero 3-form $\W_z$ over each fibre
${\cal M}_z$ as the `vertical' (fibre-wise) factor of the nowhere-zero
holomorphic 4-form over the non-compact 4-fold $M$. So, both the total
(non-compact) space $M$ and the (compact) fibres ${\cal M}_z$ are \CY.
Bertini's theorem assures us that, provided $\BM{X}$ is smooth, so is
${\cal M}^c$,  and, therefore, $\cal M$ for a generic choice of
$\phi_1$ and $\phi_2$. See ~\refs{\rBeast, \rCYCI} for more
detail on this issue.

\subsec{The spacetime metric revisited}\noindent
Next, by a theorem of Tian and Yau~\rTYnc, provided ${\cal M}_\infty$
is smooth, there is a unique
Ricci-flat K\"ahler metric in each $(1,1)$-cohomology class, for
which the associated K\"ahler form $\BM{J}$ satisfies
\eqn\eVolume{ \BM{J}^4~~
 = ~~\rd z\wedge\W_z\wedge\rd\bar{z}\wedge\ba{\W_z}
 = -\rd z\wedge\rd\bar{z}\wedge\W_z\wedge\ba{\W_z}~. }
This Ricci-flat K\"ahler metric defines, at every fibre ${\cal M}_z
=\p^{-1}(z)$ the orthogonal complement. With respect to this
decomposition, we may write
\eqn\eFourFM{ \BM{J}~
 = ~{i\over4} V(x,z)\,\rd z\wedge\rd\bar{z}~ + ~\g(x,z)~, }
where $\g(x,z)$ is a $(1,1)$-form along ${\cal M}_z$. Taking the
fourth wedge power and equating with the expression~\eVolume, we
obtain
\eqn\eGamCub{ \g^3 = {i\over V(x,z)} \W_z \wedge \ba{\W_z}~. }

Finally, note that ${\cal Z}^c = \CP1$ is not a holomorphic
subspace of $M^c$ in any natural way
 and hence does not acquire a metric by restriction. Rather, the
projection $\p:M^c \to {\cal Z}^c$ specifies ${\cal Z}^c=\CP1$
as a quotient.This leaves
open another natural possibility, using the fact that on a complex
one-dimensional space, the volume form defines a metric. Thus,
we integrate the volume form of the non-compact 4-fold $M$
over the fibre ${\cal M}_z$, point by point in ${\cal Z}$, and
obtain the induced volume form at $z\in{\cal Z}$~:
\eqn\eXXX{ \int_{{\cal M}_z} \BM{J}^4~
 = ~\int_{{\cal M}_z} i V(x,z)\rd z\wedge\rd\bar{z}\wedge\g^3(x,z)~
 = ~-\big(\int_{{\cal M}_z} \kern-.8em
           \W_z\wedge\ba{\W_z}\,\big)\, \rd z\wedge\rd\bar{z}~. }
Since $\rd z\wedge\rd\bar{z}=-i\rd^2z$, we obtain the associated
spacetime metric
\eqn\eSTM{ \rd s^2~ = -\rd t^2~ +
           ~\big(i\!\int_{{\cal M}_z} \kern-.8em
             \W_z\wedge\ba{\W_z}\,\big)\, |\rd z|^2~ + ~\rd x_3^{~2}~,}
matching exactly, and independently, our earlier result~\eSpTiMe.

\subsec{Gauge-fixing and some related properties}\noindent
{}From this bigger perspective, we can now address the gauge-fixing
which we mentioned in the previous section. From the point of view of
4-dimensional field theory in which the vacuum (as defined by the
choice of the `internal' space) varies in spacetime, we could
redefine the holomorphic 3-form $\W_z \to f(z)\W_z$, where $f(z)$ is
an arbitrary non-zero holomorphic function over the space-like
surface ${\cal Z}^c$. This indeed seems as too much freedom for
$(i\int\W\ba{\W})$ to be interpreted as a metric component. We have
remarked in the previous section that there is a (class, perhaps, of)
natural gauge(s) in which $f(z)$ is fixed so that $(i\int\W\ba{\W})$
is non-zero over all of the non-compact $\cal Z$.

Most importantly in the foregoing  analysis, $\W_z$ comes out by
definition as non-zero over all of $\cal Z$, being the 3-form factor
in the nowhere-zero holomorphic 4-form $\BM{\mit\W}$ on $M^c$;
see~\eFactOm. True, this 4-form itself is defined only up to a
rescaling $\BM{\mit\W} \to \BM{f} \BM{\mit\W}$, but this $\BM{f}$ must
be a constant over all of $M^c$! So, it merely defines a global (and
constant!) relative scale in
\eqn\eFatF{ \rd s^2~
 = ~-\rd t^2~
 + ~|\BM{f}|^2\big(i\!\int_{{\cal M}_z} \kern-.8em
                       \W_z\wedge\ba{\W_z}\,\big)\, |\rd z|^2~
 + ~\rd x_3{}^2~. }
It is of no physical consequence, for it specifies the relative
proportion of distances in the ${\cal Z}$-plane, such as the distance
between two cosmic strings, {\it vs}. some characteristic length scale
in the $(t,x_3)$-plane, where there is no structure to compare with.
In other words, the vacua with such static cosmic strings are invariant
under rescalings in the $(t,x_3)$-plane and the constant  $|\BM{f}|$
may be absorbed by these symmetries. In retrospect, we see that
$\BM{f}$ in fact equals the inverse of the overall scale $c$ in the
meromorphic section $\m$ used in~\eResOm.

It therefore follows that the metric~\eSTM\ has no physically
relevant free parameters. Also, the parameter $|\BM{f}|$ which
determines the characteristic distance between the cosmic strings is
thus unrelated to the compactification theory and is free to assume
astronomical proportions as hopefully governed by cosmological
mechanisms.

Reviewing the above analysis, we see that the formula~\eSTM\ follows
from requiring that the complex structure of ${\cal M}_z$ vary
holomorphically with $z\in{\cal Z}$ and upon declaring  a particular
point $z_\infty$ of ${\cal Z}^c$ to be the spatial infinity. Let us
remind the reader here that, in order to use the above stated theorem
of Tian and Yau~\rTYnc, $z_\infty$ must be chosen so that the
corresponding fibre ${\cal M}_\infty$ is smooth. It is not unlikely
that this condition can be relaxed to a certain extent, but it is not
clear how much. \ping

For the spacetime metric~\eSTM, the Ricci tensor equals the
Weil-Petersson-Zamolod\-chikov metric and therefore diverges
logarithmically at the cosmic string loci. Owing to Eq.~\eGamCub,
the geometry defined by the $(1,1)$-form $\BM{J}$ in~\eFourFM\ is also
singular. The arguments of \rDNAS, however, ensure the existence of a
perturbative solution to superstring theory only for smooth geometries.
We therefore need to replace the metric~\eFourFM\ on the 4-fold $M$
with a smooth one.

Here our situation is qualitatively the same as that discussed in
Ref.~\rSCS, although we cannot specify explicitly the fibre-wise
metric component better than in Eq.~\eGamCub. That is, Eq.~\eGamCub\
simply states that the volume form of $\g(x,z)$ equals the $\W_z
\wedge \ba{\W_z}$ one, normalized by $-iV(x,z)$. Unfortunately, we have
found no reason to expect any simple form for $-iV(x,z)$ in general.

Just as in Ref.~\rSCS, we temporarily re-compactify the 4-fold into
$M^c$ and recall that this is a compact K\"ahler submanifold in
${\bm X}\times \CP1$. It therefore inherits a compact K\"ahler metric
from ${\bm X}\times \CP1$ by reduction and let $J_c$ denote the
associated K\"ahler class. We form
\eqn\eXXX{ \BM{J}_\l(x,z)~
 \define ~\y(z,\bar{z}) \BM{J}(x,z) + \l J_c(x,z)~, }
where $\y(z,\bar{z})$ is a ``notch'' function which vanishes
sufficiently fast at the loci of cosmic strings\ft{Since the Ricci
tensor diverges only logarithmically at the loci of the strings, $z_i$,
near these points $\y(z)\sim|z-z_i|^\e$ suffices for however small
positive $\e$.} in $\cal Z$ and equals 1 outside a suitably small
compact region around these loci, while $\l$ is a suitably large
number. The correction term $\l J_c$ offsets the asymptotic value of
$\BM{J}_\l{}^4$ merely by an irrelevant constant since $J_c$ is a
compact metric. Near the locus of the cosmic strings, however,
$\BM{J}_\l$ is smooth and positive, as required. While the metric
associated to $\BM{J}_\l$ is not Ricci-flat, it has the advantage of
being smooth, making the arguments of Refs.~\rDNAS\ available and we
conclude that at least perturbatively, these vacua with cosmic strings
are valid solutions of superstring theory.

\subsec{1-dimensional {\it vs}. more-dimensional moduli space}\noindent
We find it necessary to point out an essential difference between the
situation encountered with a typical $b_{2,1}$-dimensional moduli
space where $b_{2,1}>1$ and the 1-dimensional special case. In many
ways, the developing theory of \CY\ modular geometry borrows from the
well understood modular geometry the torus. However, the moduli space
of the torus is 1-dimensional and care has to be taken when relying on
parallels.

In the present context, we are considering (harmonic) maps from a
punctured $\CP1$ (the puncture corresponding to the spatial infinity)
to the \CY\ moduli space. The trivial (constant) map excluded, the case
$b_{2,1}=1$ differs in a very important way from the $b_{2,1}>1$ case:
When $b_{2,1}=1$, any holomorphic map from ${\cal Z}^c$ to the moduli
space must be surjective (onto) and will in general be $m$--to--1.
Therefore, the space-like surface $\cal Z$ becomes carved up in $m$
isomorphic preimages of the moduli space, that is, fundamental domains.
Of necessity, then, the various quantities defined above (and likewise
in Ref.~\rSCS) must be identified over corresponding points in these
different copies of the fundamental domain, which places additional
non-trivial requirements on the spacetime metric structure.

In the typical case, when $b_{2,1}>1$, a holomorphic map from the
(compactified) space-like surface ${\cal Z}^c$ cannot possibly be
surjective (simply since $\dim{\cal Z}^c<b_{2,1}$). Instead, the map
will be $m$--to--1, with $m=1$ generically. Therefore, unlike when
$b_{2,1}=1$, in the typical case no additional conditions arise and
our above discussion is complete in this regard.

\vskip0pt plus2cm\penalty-250\vskip0pt plus-2cm
\newsec{The Number of Cosmic Strings and the Local Geometry}
\noindent
\subsec{Counting the strings}\noindent
There is one global property of these vacua with cosmic strings which
can be obtained easily---their total number. More precisely, there is a
notion of `charge' carried by these cosmic strings which may be defined
as follows. Assume that in the 4-fold $M$ each singular fibre $\shM_i$
at $z^\sharp_i$ has precisely one node, with $z^\sharp_i$ thus locating
the cosmic strings. By varying the defining equation(s) of $M$ some of
these strings may be brought to coincide. Suppose that at $z^\sharp_i$,
two cosmic strings from the initial setting have been brought together.
Then, the `internal' space at $z^\sharp_i$, $\shM_i$ will have two
nodes, generically at different points in $\shM_i$. We will say that
the cosmic string at $z^\sharp_i$ is doubly charged.

In a more special case, when the locations of the nodes inside $\shM_i$
are also brought to coincide, $\shM_i$ will have developed an
$A_2$-singularity (nodes are $A_1$-singularities). The cosmic string
occurring where the `internal' space develops a single $A_2$
singularity therefore will also be doubly charged, for by deformation,
that cosmic string decomposes into two singly charged ones. It is easy
to see that this provides a conserved additive charge.

The total charge of a vacuum configuration with such cosmic strings can
be obtained for the above collection of models very easily. Consider
again the compactified 4-fold $M^c$. By standard techniques, we
straightforwardly compute its Euler characteristic. Now, suppose for
a moment that $M^c$ was in fact a fibration of only smooth 3-folds
${\cal M}$ over $\CP1$, with not a single singular fibre. The Euler
characteristic would then simply be $\EU({\cal M}){\cdot}\EU(\IP^1)$.
As there exist singular fibres in $M^c$, the latter result
ought to be corrected for the (fibre-wise) singular set:
\eqn\eXXX{ \EU(M^c)~~
 = ~~ \EU({\cal M}){\cdot}\EU(\IP^1)~
 + ~\sum_{z^\sharp} \EU\big({\rm Sing}({\cal M}^\sharp_i)\big)~. }
Since each singular fibre $\shM_i$ is a conifold with isolated singular
points and $\EU({\rm point})=1$ and $\EU(\IP^1)=2$, the correction term
simply becomes the number of (fibre-wise) nodes, that is, the total
`charge':
\eqn\eStrNum{ \#_{\rm total}~~ = ~~\EU(M^c)~ - ~2\EU({\cal M})~. }
If we wish to rely only on the stronger Tian-Yau condition~\rTYnc,
which requires the fibre at infinity to be smooth, we cannot push any
of the cosmic strings off to spatial infinity. Therefore $\#_{\rm
total}$ will characterize the non-compact vacuum $M$ also. Moreover, the
Euler characteristic formula~\eStrNum\ is valid even if singularities
worse than nodes cannot be avoided owing to insufficient generality of
the model considered.

\subsec{The Local Geometry: Deficit Angle and Curvature}\noindent The
space-time metric $e^{-K}|\rd z|^2$ differs from the flat metric
$|\rd z|^2$ by a conformal factor which is $C^1$ and bounded both above
and  below in a neighborhood of each string. More explicitly, it has
the form
\eqn\eXXX{ a(\t_i,\bar\t_i) + b(\t_i,\bar\t_i)|\t_i|^2\log|\t_i|~, }
where $z_i$ is the location of a string, $\t_i \define z{-}z_i$, and
$a(0)>0$. From this it is readily calculated that there is {\it no
cone-like angular deficit at $z_i$}.

It now follows, by Gauss-Bonnet, that the curvature of the spacetime
metric integrates to $2\p\EU(\CP1)=4\p$. On the average, each singly
charged string may be thought of as contributing $4\pi/\#_{\rm total}$.
However, in this regard our situation is markedly different from that
in Ref.~\rSCS. As remarked earlier, with a (complex) 1-dimensional
space of the moduli which vary over the space-like surface $\cal Z$,
the mapping of $\cal Z$ to the moduli space is surjective and
$m$--to--1. Thus, ${\cal Z}^c$ is naturally carved up into $m$ preimages
of the moduli space. If there is one cosmic string locus as in the torus
case, integration over each one of these preimages defines the charge
of the corresponding string. Furthermore, each preimage contributes
the same amount to the total curvature $4\p$, whence the charge
(energy per unit length) of each string is a well-defined quantity and
equals $4\p/\#_{\rm total}$.

With a $b_{2,1}$-dimensional space of moduli (some of) which are
varied over the space-like surface $\cal Z$, where $b_{2,1}>1$, this
is no longer true. The mapping $t^\a$ of $\cal Z$ into the moduli
space $\cal B$ is typically injective and the various cosmic strings
are the various different intersection points of $t({\cal Z}) \subset
\cal B$ and ${\rm Sing}({\cal B})$. There is no natural way to
equipartition $\cal Z$ into $\#_{\rm total}$ regions, each of which
contains one singly charged string and contributes the same amount
to the total charge $4\p$. In this respect, the typical situation in
our case is much less determined in general.

In a concrete model, the precise locations of the strings can be
computed. Then, relying on sufficient knowledge of the periods\ft{Here,
{\bf A} and {\bf B} form a simplectic basis of $H_3$: ${\bf A}\cap{\bf
B}=\Ione$, ${\bf A}\cap{\bf A}=0$, ${\bf B}\cap{\bf B}=0$.}
\eqn\ePeriOm{ \int_{A^a}\W_z~,\qquad \int_{B_a}\W_z~, }
and so the metric component $(i\int\W\,\ba{\W})$, one may naturally
carve up $\cal Z$ into regions at the partial extrema of
$(i\int\W\,\ba{\W})$ each containing a single string. Without this
detailed model-dependent information, for which techniques are being
developed~\rPeriods, we are not aware of any further generally
valid results.

\subsec{A simple reference metric}\noindent
The general picture of the space-like surface $\cal Z$, with the
geometry obtained for it as above and with the cosmic strings passing
through it is sketched in Fig.~2.


Note, however, that this is {\it not} how the space-like surface $\cal
Z$ is embedded in the spacetime. The extrinsic curvature of $\cal Z$ in
spacetime vanishes, as should be clear from the form of the
metric~\eSTM. Also, the sketch in Fig.~2 does not show the geometry
in any fine detail; roughly, the surface $\cal Z$ appears conical
{\it around} the locus of a cosmic string, although there is no conical
deficit angle {\it at} the string.

Motivated by this, it is reasonable to introduce an auxiliary geometry,
which is indeed flat, except for a conical singularity at the locus of
each node. In order to do this, we write $z_1,\ldots,z_n$ for values
of $z$ locating the nodal varieties (accounting for multiplicities).
We set
\eqn\eXXX{ P(z)~ = ~\prod_{i=1}^n(z-z_i)~,\qquad n=\#_{\rm total}~, }
and we choose for our auxiliary flat metric
\eqn\emFlat{ \rd s^2_{\rm aux.}~~ \define ~~-\rd t^2
         + {\rd z\,\rd\bar{z}\over (P\bar P)^{1\over n}} + \rd x_3^2~.}
By straightforward computation, it is easily seen that this metric is
flat and asymptotically cylindrical at infinity, and that the deficit
angle at each $z_i$ is $2\pi\over n$. It is tempting to conjecture that
our physical metric ~\eSTM\ is a smoothing of~\emFlat, with the
curvature concentrated around the locations of the strings.
In any case, since the energy density of the cosmic strings falls off
exponentially, the metric~\emFlat\ appears to be a reasonable and useful
approximation.
To appreciate the temptation, consider the scattering problem
for~\emFlat:

The geodesic equation for~\emFlat\ is given
\eqn\eGeod{ {\rd z\over \rd t}~ = ~c\,P(z)^{1\over n}~, }
where $c$ is a complex constant in general. For large $z$, this is
asymptotic to
\eqn\eGeodAss{ {\rd z\over \rd t}~ = ~c'\,z~,}
on the understanding that the correspondence between the values of $c$
and $c'$ is $n$--to--$n$, rather than 1--1. For example, if $c$ is
real and negative, the geodesic is axial and incoming. Exceptionally,
such a geodesic may eventually hit one of the cone points. Generically,
it must come back out. When it does so, $c'$ must be replaced by
$\a c$, where $\a$ is an $n$th root of 1 with negative real part.
Geometrically, this means that the outgoing ray has acquired a
circumferential component through scattering among the strings and
where the ratio between the axial and circumferential velocities of the
outgoing ray is quantized.

\newsec{Concrete Models}\noindent
The general description of model construction in the preceding
sections may have been too vague except for the experts. We therefore
include some sample models just to illustrate the general method of
endowing families of \CY\ 3-folds with spacetime dependence.\ping

Consider firstly the one-parameter family of quintics ${\cal M}_\j$
studied recently in Ref.~\rMM:
\eqn\eXXX{ P_\j(x)~~
 \define ~~(\sum_{i=1}^5 x_i{}^5)~ - ~5\j(\prod_{i=1}^5 x_i)~. }
It possesses a $\ZZ_5^{~5}/\ZZ_5 = \ZZ_5^{~4}$ symmetry, which acts on
the $x_i$ simply by multiplying them with fifth roots of 1. The space
of complex structures of the quotient ${\cal W}_\j \define {\cal
M}_\j/\ZZ_5{}^3$ is 1-dimensional and parametrized by $\j$. To turn
this into a vacuum configuration with cosmic strings, simply allow $\j$
to vary over spacetime $X$. If, following the preceding analysis, $\j$
becomes a holomorphic function over a space-like surface ${\cal Z}
\subset X$ and we need to consider mappings from the compactified
space-like surface ${\cal Z}^c$ to the $\j$-moduli space; the moduli
space is seen as the $0\leq{\rm Arg}(\j)\leq{2\p\over5}$ wedge in the
$\j$-plane~\rMM.

At $\j=1$, ${\cal M}_\j$ develops 125 nodes and so ${\cal W}_\j$ is
likewise singular---thus $\j(z)=1$ specifies the locus of the cosmic
string (of charge 125). At $\j=\infty$, ${\cal M}_\infty$ and so also
${\cal W}_\infty$ becomes rather badly singular: ${\cal M}_\infty$ is
in fact the union of five $\CP3$'s, meeting (and singular) at ten
$\CP2$'s, which in turn have ten $\CP1$'s in common and which intersect
in five points. ${\cal W}_\infty$ is then just the $\ZZ_5{}^3$ quotient
of this and is none less singular. Since $\dim{\rm Sing}({\cal
W}_{\infty})>0$, the total space $M^c$ of this $\j$-family of \CY\
3-folds is singular at $\j=\infty$. To construct a smooth non-compact
$M$ from $M^c$, we therefore must excise the singular fibre and then
the theorem of Tian and Yau~\rTYnc\ does not guarantee the existence of
a Ricci-flat metric $M$; this then remains an open question.
Nevertheless, the metric on the 4-dimensional spacetime $X$ is still
determined by ~\eSTM. Hoping that Ricci-flatness can eventually be
proved, and because detailed computations are available from Ref.~\rMM,
we have chosen to include this example here, and indeed to refer to it
throughout the paper. \ping

Consider next another family of quintics in \CP4, one that has been
studied in Refs.~\refs{\rRoll, \rWeb}~:
\eqn\eXXX{ {\mit\D}_t(x)~~
 \define ~~\big(\,S(x)P(x)-Q(x)R(x)\,\big)~ - ~t\, T(x)~~ = ~0~, }
where $S, Q, P, Q$ are generic polynomials with $\deg S=\deg Q=q$,
$\deg P=\deg R=(5{-}q)$ and $T(x)$ is a generic quintic in \CP4. At
$t=0$, ${\mit\D}_0$ is singular and has precisely $q^2(5{-}q)^2$
nodes. This is easily seen since
\eqn\eXXX{ \rd{\mit\D}_0(x)~~
 = ~~\rd S\,P + S\,\rd P - \rd Q\,R - Q\,\rd R }
vanishes when $S=P=Q=R=0$. This being four (generically) independent
conditions in \CP4, they will be satisfied at isolated points. The
number of these points follows by B\'ezout's theorem and equals the
product of degrees. Near $t=0$, the quintics ${\mit\D}_t=0$ are smooth,
but eventually, as we let $t$ vary linearly over a ${\cal Z}^c= \IP^1$,
there will be other singular loci. Interperting $t$ as a spacetime
field varying holomorphically over the space-like surface $\cal Z$, we
obtain a vacuum configuration with cosmic strings.

To use the formula~\eStrNum, we need that the Euler characteristics
of smooth quintics in \CP4 is $-200$. Likewise, the Euler
characteristic of 4-folds of bi-degree $(5,1)$ in $\CP4
\times \CP1$ (corresponding to a linear fibration over $\cal Z$) is
easily found to be $+880$, whence the total cosmic string charge
becomes $\#_{\rm total}=1280$. The precise location, in $\CP1_t$, of
the remaining 1264 strings (several subsets of which may be coinciding
in location) depends on the specific choice of the polynomials
$S,P,Q,R$ and $T$. Since $b_{2,1}=101$ for smooth quintics in \CP4,
we see that there is a wealth of possible vacuum configurations with
(static) cosmic strings. As long as the fibration is linear as above,
$\#_{\rm total}$ remains at 1280.
\ping

The well known Tian-Yau family of \CY\ 3-folds is defined as the common
zero-set of
\eqn\eXXX{{\eqalign{
 \sum_{i=0}^3 x_i{}^3~ -
          ~3 \sum_{i,j,k=0}^3 \a_{ijk} x_i x_j x_k~ &= ~0~, \cr
 \sum_{i=0}^3 y_i{}^3~ -
          ~3 \sum_{i,j,k=0}^3 \b_{ijk} y_i y_j y_k~ &= ~0~, \cr
 \sum_{i=0}^3 x_i y_i                     ~ &= ~0~. \cr}}}
For suitable choices of the constants $\a$ and $\b$, these 3-folds,
${\cal M}^0_{\a,\b}$, are smooth and admit a free $\ZZ_3$ action and
the quotients ${\cal M}_{\a,\b} \define {\cal M}^0_{\a,\b}/\ZZ_3$ are
smooth \CY\ 3-folds with $\EU=-6$. We may then make any of $\a,\b$ into
a $\cal Z$-dependent spacetime field and obtain thereby a vacuum
configuration with cosmic strings. Again, making any of $\a,\b$ into a
linear function over $\cal Z$ yields a linear fibration. Temporarily
compactifying the space-like surface into ${\cal Z}^c=\CP1$, we have
just constructed a compact 4-fold embedded as the common zero set of a
system of three equations in $\CP3 \times \CP3 \times \CP1$ and of
tri-degree $(3,0,1)$, $(0,3,0)$ and $(1,1,0)$. The Euler characteristic
of this 4-fold is +135, whence $\#_{\rm total} = 135-2(-18) = 171$. As
the whole family admits the $\ZZ_3$ symmetry, by passing to the
$\ZZ_3$-quotient, we obtain a 3-generation vacuum configuration with 57
(possibly coinciding) cosmic strings.\ping

The simplicity of these constructions is we hope clear. Of course,
more detailed information on the spacetime geometry, such as an exact
expression for the spacetime metric throughout $\cal Z$, depends on
the knowledge of periods~\ePeriOm. Certain preliminary and general
results are already known in this respect and may soon become
routine calculation~\rPeriods.

\newsec{Some Unusual Global Properties}\noindent
We have so far seen several unusual properties of the cosmic strings
that arise in the present context. Below we discuss another
feature which is caused by certain delicate relations between complex
structure moduli and (complexified) K\"ahler class moduli at certain
singular \CY\ 3-folds.

Consider the family of \CY\ 3-folds embedded in \CP5 by means of the
two equations
\eqn\eXXX{ {\cal M}_\g~:\qquad \left\{
    {\eqalign{ \sum_{i=0}^5 x_i{}^4~        &= ~0~, \cr
               x_2 x_3 - x_4 x_5 +\g\,Q(x)~ &= ~0~, \cr}} \right.}
where $Q(x)$ is a generic quadric in \CP5. At $\g=0$, ${\cal M}_0$
becomes singular and has four nodes, located in \CP5 at
$(1,\x,0,0,0,0)$, where $\x^4=-1$. Near $\g=0$, ${\cal M}_{\g\ne0}$
are smooth and a simple local calculation~\rRoll\ finds four small
3-spheres $S^3_\x \subset {\cal M}_{\g\ne0}$, each of which collapses
troduced if the singular fibre is smoothed by small resolutions
instead of deforming~\refs{\rBeast, \rRoll, \rWeb}. This number is
also known as the `deficit'~\rClem.

Another unusual property of these fibrations is the fact that the whole
family of \CY\ 3-folds over the space-like surface $\cal Z$ is given
globally by means of some system of equations as above. As defined so
far, these constructions provide static vacuum configurations. However,
if we try to change the family so as to relocate some of thhis remark is to
note that not all nodes in a fibre
are independently smoothed by deformations in fibrations of the kind
we are examining.  In fact, the number of relations among these
deformations by smoothing equals the number of vanishing cycles which
are introduced if the singular fibre is smoothed by small resolutions
instead of deforming~\refs{\rBeast, \rRoll, \rWeb}. This number is
also known as the `deficit'~\rClem.

Another unusual property of these fibrations is the fact that the whole
family of \CY\ 3-folds over the space-like surface $\cal Z$ is given
globally by means of some system of equations as above. As defined so
far, these constructions provide static vacuum configurations. However,
if we try to change the family so as to relocate some of the cosmic
strings in the $\cal Z$-plane---this will inevitably have global
consequences and likely also move some strings well away from the
strings which we were initially relocating. This may appear unsettling,
as it would suggest some sort of action at distance, which seems to run
against the usual ideas of locality. The resolution is of course in the
fact that we have from the outset restricted to static and
$x_3$-independent vacua and this ought to be considered merely as an
approximation to more realistic vacua in which the cosmic strings would
have full spacetime dependence and then presumably a corresponding
local dynamics. We now turn to some preliminary discussion of such
ideas.

\newsec{The Full (10-Dimensional) Picture}\noindent
In full generality, the moduli of the `internal' space should have full
spacetime dependence and we need to solve the equations of motion for
the moduli~\eEqMotn\ and the Einstein equations~\eEqEins\ for the
metric. If one performed a Wick-rotation into imaginary time, spacetime
becomes Euclidean and if we adopt the Ansatz that it also admits a
complex structure, we are able to proceed in close analogy with the
static case.

We start, as before, with a (complex) 4-fold $\bm X$ whose
anti-canonical bundle is non-trivial and generated by global sections.
We now assume that $K_{\bm X}\!\con$ has at least three linearly
independent global holomorphic  sections, which constitute a basis of
a 3-dimensional space of linear combinations
$U=\{\x_1\f_1+\x_2\f_2+\x_3\f_3\}$, which we projectivize into a
$\IP^2_\x$-family of sections of $K_{\bm X}\!\con$. The projective
space $\IP^2_\x=X^c$ will be interpreted as the compactified and
Wick-rotated spacetime, endowed moreover with a complex structure.
A {\it generic} section $\F(x,\x)$ from $U$ then defines a smooth \CY\
3-fold ${\cal M}_\x$ lying, therefore, over a generic point of
$\x\in X^c$. The total space of this (projective) family is a
non-singular complex 5-fold ${\bm M}^c_X$, with the projection
\eqn\eXXX{ \pi: {\bm M}^c_X \longrightarrow \IP^2_\x = X^c~, }
of which the generic fibre is a smooth Calabi-Yau 3-fold. That is, the
compact 5-fold ${\bm M}^c_X$ has the structure of a fibre space, with
$\IP^2_\x$ as the base an \CY\ 3-folds ${\cal M}_\x$ defined for each
$\x\in X^c=\IP^2_\x$ by $\F(x,\x)=0$.

\subsec{The singular set}\noindent
Again, the r\^ole of the world sheet of the cosmic string(s) is played
by the singular locus in ${\bm M}^c_X$ of the projection
$\pi:{\bm M}^c_X \to \IP^2_\x=X$, which in this case is an algebraic
curve (Riemann surface). We will consider this first for the compact
5-fold ${\bm M}_{\cal Y}^c$ and later interpret its intersection with a
non-compact total 10-dimensional spacetime \BM{M}.

Let ${\bm P}^\sharp_{\bm X}$ denote the projectivized space of global
sections of $K_{\bm X}\!\con$ for which the zero locus is singular.
Recall that it is a hypersurface in ${\bm P}_{\bm X}$, the projective
space of all global sections of $K_{\bm X}\!\con$. The degree $d$ of
this hypersurface is given by
\eqn\eDegNum{ d ~=~ \EU(M^c_{\cal Z}) - 2\EU({\cal M})~, }
where ${\cal Z}=\IP^1$ is a linear subspace of ${\bm P}_{\bm X}$,
$M^c_{\cal Z}$ is a $\IP^1$-family of \CY\ 3-folds, defined in
Eq.~\eDefEq, while ${\cal M}$ is the generic \CY\ 3-fold defined as the
zero locus of a non-singular global section of $K_{\bm X}\!\con$. This
is simply a reinterpretation of the computation of the number of
strings for the static case, Eq.~\eStrNum.

Now, by identifying $\IP^2_\x \subset {\bm P_X}$ with the compactified
(real 4-dimensional) spacetime $X^c$, the latter has been mapped into
the projective space of sections of $K_{\bm X}\!\con$, which is a
moduli space for \CY\ 3-folds ${\cal M}_\x\subset{\bm X}$, up to
reparametrizations\ft{Please note that $X$ denotes the (real)
4-dimensional spacetime which is here Wick-rotated, while the boldface
symbol $\bm X$ refers to the complex 4-fold in which the `internal'
\CY\ 3-fold is embedded.}. As ${\bm P}^\sharp_{\bm X}$ is a
hypersurface in \BM{P_X}, the image of $X^c$ in \BM{P_X} is bound to
intersect ${\bm P}^\sharp_{\bm X}$ in a complex 1-dimensional space.
The inverse image of this is a projective plane curve ${\cal S}\subset
X^c$ of degree $d$.

In other words, at the spacetime points $\x\in{\cal S}\subset X^c$ the
`internal' space ${\cal M}_\x$ develops a node and fibres with two nodes
correspond to transverse self intersections of $\cal S$. For a generic
choice of $\IP^2_\x\subset{\bm P_X}$ and if the number of global
holomorphic sections is sufficiently big, fibres with more than two
nodes or with worse singularities will not occur. It follows that the
set $\Tw{\cal S}\subset{\bm M}^c_X$ of singular points of the mapping
$\pi$ is a non-singular curve, in fact the desingularization of $\cal
S$.

Remembering that the Euler characteristic of each nodal fibre exceeds
the Euler characteristic of a smooth fibre by the number of nodes on
the fibre, we obtain the equation \big(cf.~\eStrNum\ and~\eDegNum\big)
\eqn\eGenNum{ \EU(\Tw{\cal S})~ = ~\EU({\bm M}_X^c)~-~3\EU({\cal M})~,}
since $\EU(\IP^2)=3$; the number of handles on the cosmic string world
sheet is then $g=1-\inv2\EU(\Tw{\cal S})$. There is good technology for
computing the Euler characteristics of smooth 3-folds ${\cal M}$, the
total space of their linear $\IP^1_z$-families $M^c_{\cal Z}$, and the
total space of their linear $\IP^2_\x$-families $\BM{M}_X^c$~\rBeast.
Therefore, Eqs.~\eDegNum\ and~\eGenNum\ yield the degree $d$ and the
genus $g=1-\inv2\EU(\Tw{\cal S})$ of the world sheet of the cosmic
string.

The number $n^\sharp$ of self-intersections of ${\cal S}$ (nodes of the
cosmic string world sheet) is then determined by the equation
\eqn\eSIntNum{n^\sharp~ =
              ~{1\over2} \big[ \EU(\Tw{\cal S})~-~(3d-d^2) \big]~.}
This follows from the fact that a non-singular plane
curve of degree $d$ has Euler characteristic $3d-d^2$ and that the
Euler characteristic of the desingularization of a plane curve of the
same degree with transverse self intersections exceeds that of a
non-singular plane curve by twice the number of self-intersections.

\subsec{Decompactification and Metric}\noindent
In order to decompactify ${\bm M}^c_X$ and obtain a non-compact
5-fold ${\bm M}_X$ with a Ricci-flat metric, we choose a
projective line $\IP^1_\infty \subset \IP^2$, recalling that
$\big(\IP^2_\x-\IP^1_\infty\big) = \IC^2 \approx \IR^4$. For a generic
choice, the total space of the family of \CY\ 3-fold fibres over this
$\IP^1_\infty$, $\pi^{-1}(\CP1_\infty)$, is a non-singular hypersurface
in ${\bm M}^c_X$. This hypersurface is the locus of the pole of an
otherwise holomorphic and non-vanishing 5-form
\eqn\eResOM{ {\eqalign{ \BM{\mit\W}(x,\x)~
 &\define ~~{\rm Res}_{(x,\x)}
   \Big[\,{c\over C(\x_1,\x_2,\x_3)} {\w\over\F(x,\x)} \,\Big]~, \cr
 &= ~~\lim_{\G\to(x,\x)}
       \oint_{\G} {c\over C(\x_1,\x_2,\x_3)} {\w\over\F(x,\x)}~,
        \qquad (x,\x) \in {\bm M}^c_X~,                          \cr}}}
where $C(\x)=L^\D(\x)$, with $\D=3-\deg_\x\F(x,\x)$ and $L(\x)$ is
linear in $(\x_1,\x_2,\x_3)$. This is of course very much like
${\cal M}_{z_\infty}$ over spatial infinity $z_\infty\in\IP^1_z$, which
was the locus of the pole of the otherwise holomorphic and
nowhere-zero 4-form \BM{\mit\W}~\eResOm. We note also that a generic
choice of $\CP1_\infty$ will meet ${\cal S}$, the set of singular
points of the singular fibres,  transversely in $d$ points on distinct
fibres.

Now we can invoke the theorem of Yau and Tian to deduce the existence
of a Ricci-flat metric on
${\bm M}_X = {\bm M}^c_X - \pi^{-1}(\CP1_\infty)$.
\ping

We take our four dimensional spacetime to be $X=(\IP^2_\x-\IP^1_\infty)
\approx\IR^4$. The metric on ${\bm M}$ induces both a horizontal
subspace wherever $\pi$ is non-singular and a metric on each fibre.
By integrating the horizontal metric over the fibres, we produce a
metric at each point of $X$ which is not a singular value of $\pi$.
Unfortunately, there is no reason to believe either that this metric
is K\"ahler or that the logarithm of its volume form is a K\"ahler
potential for the Weil-Petersson-Zamolodchikov metric.

The fact that $c_1(\IP^2)=3J_{\IP^2}$, allows another construction.
With $\F(x,\x)$ linear in $\x$, the quadratic function $C(\x)$ in
Eq.~\eResOM\ could be replaced by a product of distinct linear factors,
say $L_1\,L_2$, whereby the singular locus becomes the union of two
\CP1's meeting  transversely in a single point. This offers an
interesting possibility for defining Wick-rotation back to Lorentzian
spacetime, by choosing $t=\log|{L_1\over L_2}|$ as a time coordinate.
Ignoring (or perhaps blowing up) the  point $L_1=L_2=0$, we may
identify  $L_1=0$ as the light-cone 2-sphere at the infinite past and
$L_2=0$ as the light-cone 2-sphere at infinite future. For generic
choices, each of these two light-cone 2-spheres will have $d$ points
corresponding to singular fibres, and $\cal S$ will exemplify the by
now standard picture of the world sheet of an interacting string as
advanced by Mandelstam~{\rMan} and others, in which $d$ cylinders
emerge from the remote past, they join and diverge finitely many times,
and $d$ cylinders continue into the remote future. The authors hope to
explore this idea in more detail in a future publication. One
difficulty to be overcome is that fact that $\rd t=0$ at finitely many
points of ${\bm M}_X$, all of which are on the string.
\ping

We close with the following remark. At least in the main case under
study here, when the moduli fields vary only over a space-like surface,
the spacetime metric~\eSTM\ differs from the flat
metric by a conformal factor which in fact is the K\"ahler potential of
the Weil-Petersson-Zamolodchikov metric, in a suitable gauge:
\eqn\eXXX{ g_{z\bar z} ~=~ e^{-K}~. }
On the other hand, the image of the 4-dimensional spacetime in the
moduli space acquires a metric by restriction, and so there is another
`natural' metric: the pull-back of the Weil-Petersson-Zamolodchikov
metric:
\eqn\eXXX{ G_{z\bar z} ~=~ \vd_z \vd_{\bar z} e^{-K}~. }
Besides a possible application of the above analysis in cosmology, the
relation between $g_{z\bar z}$ and $G_{z\bar z}$ enables one to use
the spacetime variable vacua as a laboratory for studying the
Weil-Petersson-Zamolodchikov geometry of the moduli space. Of course,
results of this modular geometry have already been used in our study
of the spacetime variable vacua and further results will again acquire
their cosmological interpretation through these cosmic string models.
It is our hope that perhaps this interpretation will bring about a
cross-disciplinary communication in the other direction also.

 %
\vfill
{\ninepoint\noindent{\bf Acknowledgments}:
We are grateful to B.~Greene, P.~Steinhart, C.~Vafa, E.~Witten and
S.-T.~Yau for valuable discussions in the various stages of this
project. T.H.\ was supported in part by the DOE grant DE-FG02-88ER-25065
and would also like to thank the Department of Mathematics of the
National Tsing-Hua University at Hsinchu, Taiwan, for the warm
hospitality during the time when part of this research was completed.}

\vfill\eject

\listrefs
 %
 %
\bye